\documentclass[fleqn,twoside]{article}
\usepackage{espcrc2}

\usepackage{graphicx}
\usepackage[figuresright]{rotating}

\newcommand{\AmS}{{\protect\the\textfont2
  A\kern-.1667em\lower.5ex\hbox{M}\kern-.125emS}}

\title{Renormalized QCD-inspired model for the pion and mesons}

\author{T. Frederico\address{Dep. de F\'\i sica, Instituto Tecnol\'ogico
de Aeron\'autica, Centro T\'ecnico Aeroespacial, \\ 12.228-900,
S. Jos\'e dos Campos, SP, Brazil 
},
M. Frewer  
\address[MPI]{ Max-Planck Institut f\"ur Kernphysik,
D-69029 Heidelberg, Germany}
and Hans-Christian Pauli \addressmark[MPI]
}
       
\begin{document}

\begin{abstract}
    We apply the subtraction method  to an effective QCD-inspired model,
    which includes the Coulomb plus a zero-range hyperfine interactions,
    to define a renormalized Hamiltonian for mesons.  
    The spectrum of the renormalized Hamiltonian agrees with the one 
    obtained with a smeared hyperfine interaction.  
    The masses of the low-lying  pseudo scalar and vector mesons are
    reasonably described within the model.
\vspace{1pc}
\end{abstract}

\maketitle

\section{Introduction}

We address to the effective mass operator equation of the
$\uparrow\downarrow$-model for the $q\overline q$ Light-Front
Fock state component of the meson with mass $M$: 
\begin{eqnarray}
\left[ M^2-4m^2-4k^2\right]\varphi (\vec k)=\int d\vec p\; U(\vec k,{\vec p} )
\varphi ({\vec p}), \!\!\!
\label{pauleq}
\end{eqnarray}
with the kernel
$$ U(\vec k,\vec p )=-\frac{4\alpha}{3\pi^2m}
\left[ \frac{2m^2}{(\vec k-\vec p )^2}+1 \right].$$
Equation (\ref{pauleq}) is mathematically not defined.

The aim of this paper is to give Eq.(\ref{pauleq}) a physical meaning by
renormalization, i.e., by applying to it  the
``subtraction method'', a renormalization scheme for nonrelativistic 
quantum mechanics of singular interactions developed earlier \cite{t1}.

This is an interesting problem because Eq.(\ref{pauleq}) 
as proposed in  Ref.\cite{pauli4}, the $\uparrow\downarrow$-model,
is an effective Hamiltonian  derived from Quantum Chromodynamics, meant
to describe the lowest Fock-state component of the Light-Front meson 
wave-function. 
It has been applied with reasonable success to the low-lying 
pseudo-scalar and vector mesons, by using a different renormalization
scheme namely by regularization and subsequent renormalization
\cite{pauli4}.  
To the authors knowledge, fortunately nobody tried at that time 
in the past to solve the hyperfine
spliting in the hydrogen atom using the Schroedinger equation directly.
We are thus in the unique position to compare two drastically different 
schemes, both conceptually and numerically, and verify that they agree.
This strong statement stays at the very basis of renormalization
ideas, that no matter the intermediate steps one performs to  
 mathematically define the initial equation (\ref{pauleq}), 
after renormalization all them produce the same physics.   

\section{Notation}
For the purpose of presenting the subtraction method of Ref.\cite{t1},
we introduce the notation below, and  allow  as well
different quark masses and  the relativistic phase space.
\begin{eqnarray}
\left( M_0^2+V+V^\delta \right)|\varphi >= M^2 |\varphi >\ ,
\label{mass2}
\end{eqnarray}
where the free mass operator of the 
quarks with masses $m_1$ and and $m_2$ is $M_0=E_1+E_2$. The individual
energies are $E_i=\sqrt{m_i^2+k^2}$ ($i$=1,2) and $k\equiv|\vec k|$.
The Coulomb-like effective potential is $V$ and 
the hyperfine singular  interaction is $V^\delta$, which 
in the non-relativistic limit is the Dirac-delta. 
The matrix elements of these operators in momentum representation 
are given by:
\begin{eqnarray}
<\vec k|V|\vec{p }>=-\frac{4m_s}{3\pi^2}<\vec k|\chi>\frac{\alpha}{
Q^2}<\chi|\vec{p }> , \; \; 
\label{mecoul}
 \\
<\vec k|V^\delta|\vec{p }>=
   <\vec k|\chi> \frac{\lambda}{m_r}<\chi|\vec{p }> ,
\; \; \; \; \; \; \; \; \; \; \;  
\label{mesing} \end{eqnarray}   
where the total and reduced mass are denoted by 
$ m_s=m_1+m_2$ and $m_r=m_1m_2/m_s$, respectively.
$\lambda$ is the bare strength of the Dirac-delta 
hyperfine interaction.
The mean four-momentum transfer is  taken as 
$Q^2=(\vec k-\vec p  )^2$.
The phase-space dimensionless function is 
\begin{eqnarray}
\frac{1}{A(k)}=m_r\frac{E_1 +E_2}{E_1E_2} .
\label{phsp}
\end{eqnarray}
The form-factor is 
$\chi(k)=<\vec k|\chi>=1/\sqrt{A(k)}$.

\section{Example: Dirac-Delta potential}

Many authors in the past have renormalized the Schroedinger equation
with contact interaction (see e.g. \cite{glazek}). Here we want
just to supply the essence of the ``subtraction method'' 
of Ref.\cite{t1}. Let us solve Eq.(\ref{pauleq}) with only 
\begin{eqnarray}
 U(\vec k,\vec p  )=\lambda,
\label{delta}
\end{eqnarray} 
which, as said, makes Eq.(\ref{pauleq}) not well defined.

The scattering amplitude for the potential (\ref{delta}) is  the geometrical
series 
\begin{eqnarray}
\tau (M^2)=\left[\lambda^{-1} - I(M^2)\right]^{-1},
\nonumber
\end{eqnarray}
 solution of the scattering equation
\begin{eqnarray} 
T(M^2)= V+VG^{(+)}_0(M^2)T(M^2),
\nonumber
\end{eqnarray}
for the scattering state of mass $M$.
The Green's function of the free mass operator equation with outgoing
wave boundary condition is
\begin{eqnarray}
G^{(+)}_0(M^2)=\left[M^2-M^2_0+i\varepsilon\right]^{-1} .
\nonumber
\end{eqnarray}
The function 
\begin{eqnarray}
I(M^2)=\int d\vec k \frac{1}{M^2-4m^2-4k^2+i\varepsilon}
\nonumber
\end{eqnarray}
diverges linearly! This is the mathematical problem in Eq.(\ref{pauleq}).

How to give meaning to $\tau(M^2)$? We use the renormalization idea. Suppose
$\tau(\mu^2)$ is known from  experiment, then we rewrite $\tau(M^2)$
using this piece of data:
\begin{eqnarray}
\tau (M^2)=\left[\tau^{-1}(\mu^2)+I(\mu^2) - I(M^2)\right]^{-1}
\end{eqnarray}   
and know the subtraction of the divergence appears! A closer look to
\begin{eqnarray}
I(\mu^2)-I(M^2)=(M^2-\mu^2)\int d\vec k 
\nonumber
\end{eqnarray}
$$\times\frac{1}{(\mu^2-4m^2-4k^2+i\varepsilon)
 (M^2-4m^2-4k^2+i\varepsilon)} $$
shows that it is finite with  $\mu$ being the subtraction point! 
This is the essence of the ``subtraction method'' of Ref.\cite{t1}.

The  renormalized form of the potential (\ref{delta}), 
$V^\delta_{\cal R}$, has the bare strength 
 written as a function of the renormalized
one $\lambda_{\cal R}(\mu^2)=\tau(\mu^2)$
\begin{eqnarray}
 \lambda=\frac{1}{1+ \lambda_{\cal R}(\mu^2)I(\mu^2)}
\lambda_{\cal R}(\mu^2), 
\label{lren}
\end{eqnarray}
in which the physical input and the counter terms that subtract 
all the infinites in the scattering matrix at the mass scale $\mu$
are present\cite{t1}. 

\section{ Renormalized Model}

The ``subtraction method'' exemplified in sec.3 is applied
to the effective model defined by the mass operator of Eq.(\ref{mass2}).
The scattering matrix comes from the solution of the
scattering equation with the renormalized potential\cite{t1,tp}
\begin{eqnarray}
T_{\cal R}(M^2)= V_{\cal R}+ V_{\cal R}G^{(+)}_0(M^2)T_{\cal R}(M^2),
\label{trenv}
\end{eqnarray} 
where $V_{\cal R}=V+V_{\cal R}^\delta$.
In finding the bound state, one could as well diagonalize the
 mass operator $M_0^2+V_{\cal R}$ \cite{t1}.

The renormalized Dirac-delta interaction  is written  
formally as below\cite{t1}:
\begin{eqnarray}
V^\delta_{\cal R} =
\left[1+T^\delta_{\cal R}(\mu^2) G^{(+)}_0(\mu^2)\right]^{-1}
T^\delta _{\cal R}(\mu^2) ,
\label{vfren1}
\end{eqnarray}
where $T^\delta_{\cal R}(\mu^2)$ is the renormalized T-matrix of 
the Dirac-delta interaction, with matrix elements given by
$$\; \; <\vec p|T^\delta_{\cal R}(\mu^2)|\vec q>=
\chi(p)\lambda_{\cal R}(\mu^2) 
\chi(q).$$

The solution of Eq. (\ref{trenv}) is \cite{tp}:
\begin{eqnarray}
T_{\cal R}(M^2)=T^V(M^2)+|F>t_{\cal R}(M^2)<\overline F| ,
\label{tv}
\end{eqnarray}
where
$$|F>=\left( 1+T^V(M^2)G^{(+)}_0(M^2)\right) |\chi>  , $$
$$<\overline F|=<\chi|\left(G^{(+)}_0(M^2)T^V(M^2)+1\right) , $$
$$ t^{-1}_{\cal R}(M^ 2)=\lambda_{\cal R}^{-1}(\mu^2)
-{\cal G}(M)+{\cal G}_0(\mu)  , $$ 
$$
{\cal G}(M)=< \chi |\left[M^2-M^2_0-V+i\varepsilon\right]^{-1} |\chi>, $$ 
and
$${\cal G}_0(M)=< \chi|\left[M^2-M^2_0+i\varepsilon\right]^{-1}|\chi>.$$
The regular potential T-matrix, $T^V(M^2)$, is the solution of 
the scattering equation (\ref{trenv}) for  $V$.

The scattering equation
with the renormalized interaction appears in a subtracted form \cite{t1},
in which all the divergent momentum integrals are explicitly removed:
\begin{eqnarray}
&& T_{\cal R}(M^2)=T_{\cal R}(\mu^2)\mbox{\Large $\left[   \right.$}1
\nonumber \\
&& +\left. \left( G^{(+)}_0(M^2)- G^{(+)}_0(\mu^2)
   \right)T_{\cal R}(M^2)\right] .
 \label{tvren7} 
\end{eqnarray} 
For a regular potential Eq.(\ref{tvren7}) is  completely 
equivalent to the traditional Lippman-Schwinger scattering equation.

We use the renormalization condition that at the pion mass, $M=m_\pi$, the
T-matrix, for $m_1=m_2=m_u=m_d$ has a bound-state pole, 
where $t^{-1}_{\cal R}(m^2_\pi)=0$. The
 choice $\mu=m_\pi$  implies that
$\lambda_{\cal R}^{-1}(m_\pi^2)= {\cal G}(m_\pi)-{\cal G}_0(m_\pi)$.

The physics described by the theory does not dependent on
the arbitrary renormalization point, this 
 imposes $\frac{d}{d\mu^2}V^\delta_{\cal R}=0$ and
qualifies the interaction as the fixed-point of this equation.
The inhomogeneous term of Eq.(\ref{tvren7}) runs 
 as the subtraction point moves, according to the Callan-Symanzik equation
\begin{eqnarray}
\frac{d}{d \mu^2} T_{\cal R}(\mu^2)=-T_{\cal R}(\mu^2)
G^{(+)}_0(\mu^2)^2T_{\cal R}(\mu^2) .
\label{tren7}
\end{eqnarray} 
The renormalized T-matrix (\ref{tv}) is invariant 
under the change of $\mu$ to $\mu'$, and thus
$\frac{d}{d\mu^2}t_{\cal R}(M^2)=0$, and the renormalized strength runs as
 according to 
$\lambda_{\cal R}^{-1}(\mu'^2)= \lambda_{\cal R}^{-1}(\mu^2)
+{\cal G}(\mu)
-{\cal G}(\mu ')-{\cal G}_0(\mu)
+{\cal G}_0(\mu ')$.

The bare strength is obtained by equating
Eq.(\ref{mesing}) to Eq.(\ref{vfren1}), and  using 
 $\lambda_{\cal R}(m_\pi^2)$, one finds
$\lambda_{bare}=m^\pi_r{\cal G}(m_\pi)^{-1}$,
with the reduced mass $m_r^\pi=1/2$ and ${\cal G}(m_\pi)$
 calculated for 
$m_1=m_2=m_u=m_{\overline d}$.  With this, the pole of the 
T-matrix (\ref{tv}) at the bound-state mass, $M_b$, is given by
\begin{eqnarray}
t^{-1}_{\cal R}(M_b^2)=\frac{m_r}{m_r^\pi}{\cal G}(m_\pi)
-{\cal G}(M_b)=0 \ , 
\label{trenv23}
\end{eqnarray}
for  s-wave states with any quark mass. 

\section{Comparing Renormalization Schemes}

Here we compare the  results obtained with the Yukawa form of the smeared 
Dirac-delta interaction\cite{frewer}, Eq.(\ref{mass1c}), 
and the ``subtracion method'', Eq.(\ref{p23r}), with $A(k)=1$. 
Using Eq.(\ref{mass1c}), the pion mass, $m_\pi$, of 140 MeV and 
the first excited s-wave state mass, $m^*_\pi$, of 768 MeV  were fitted 
with the parameters $\alpha=$ 0.763 and $\eta$= 1148MeV see \cite{frewer}, 
for $m=$ 406 MeV. The solution of Eq.(\ref{p23r}), for $m_\pi=$ 140 MeV and
$\alpha=$ 0.763 is $m^*_\pi=$ 766 MeV, in remarkable 
agreement with the previous result.

Both renormalization methods have the same physical inputs. In one set of 
calculations, $\alpha$ was varied, with a fixed $m_\pi$= 140 MeV.
In the other set of calculations, $m^*_\pi$= 768 MeV was
kept fixed. 
In the model of Eq.(\ref{mass1c}) the value of $\eta$ was fitted to
the $m_\pi$ or $m^*_\pi$ for a given $\alpha$.
 
In figure 1, the results of $m^*_\pi$ as a function of $\alpha$ 
for the two renormalization methods are shown.
The agreement between the ``subtraction method''
and the smeared delta renormalization method is within few percent,
which we relate to the rather drastically different methods. 
The values of $\eta$ 
for $\alpha$ going to zero increase towards infinite, to keep
the ground state at the pion mass, while $m^*_\pi$
tends to the scattering threshold at 812 MeV. For $\alpha$ increasing
the values of $\eta$ decreases to keep $m_\pi$ fixed, and the 
$m^*_\pi$ which is Coulomb dominated, 
has to  decrease, as we observe in figure 1. 
The effect of the relativistic phase-space, Eq.(\ref{phsp}),
has been studied in Ref.(\cite{tp}) and it is of the order 
of only few percents.

The results for $m_\pi$  as a function of 
$\alpha$ for $m^*_\pi=$ 768 MeV, are presented in figure 2. 
The threshold for zero pion mass occurs for 
$\alpha$ with the value about 0.75.  
The value of $m_\pi$ increases with $\alpha$, corresponding to a decreasing
binding energy, which means that the intensity of the short-range 
interaction, that dominates the ground state, diminuishes. In fact
to keep constant $m^*_\pi$, as the effective Coulomb interaction
increases it demands a weaken  short-range interaction. The calculation
of $m_\pi$ with Eq.(\ref{mass1c}) does not go beyond $\alpha=0.97$ 
because $\eta$ vanishes and the mass of 768 MeV of the excited state is 
reproduced with the effective Coulomb interaction. The 
``subtraction method'' does not present the same limitation.

\begin{figure}[t] 
\begin{center}\scalebox{0.4}{\includegraphics{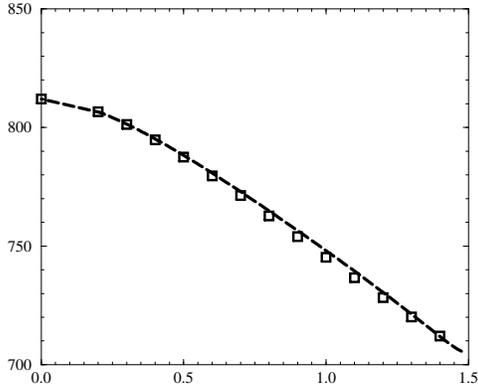}}
\end{center}
\vspace {- 2cm}
\caption{\label{fig1}
The mass $m^*_\pi$(MeV) is plotted versus $\alpha$   
for a fixed $m_\pi=$ 140 MeV. 
The dashed curve gives 
results from Eq.(\protect{\ref{mass1c}}),  the empty boxes
from Eq.(\protect{\ref{p23r}}) with $A(k)=1$.
}\end{figure}

\begin{figure}[t] 
\begin{center}\scalebox{0.4}{\includegraphics{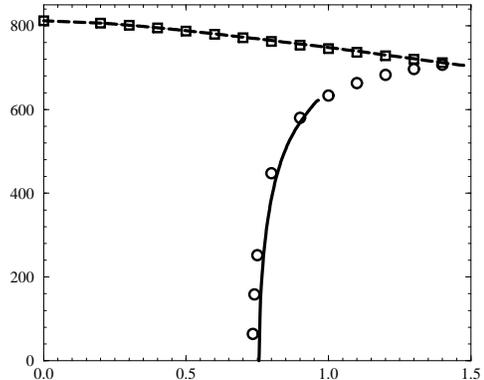}}
\end{center}
\vspace{-2 cm}
\caption{\label{fig2}
The mass $m_\pi$(MeV) is plotted versus $\alpha$   
 for a fixed $m^*_\pi=$ 768 MeV. The solid curve gives 
results from Eq.(\protect{\ref{mass1c}}),  the empty circles
from Eq.(\protect{\ref{p23r}}) with $A(k)=1$. 
The two upper curves are the ones of fig. 1.
}\end{figure}

\section{Effective Meson Model}

Now, we use the ``subtraction method'' applied to the 
$\uparrow\downarrow$-model to calculate the mass gap between the
ground states of the pseudo-scalar and vector mesons, corresponding
to ($\pi ,\rho$)[139,768], ($K^\pm ,K^*$)[494,892], 
($D^0,D^{*0}$)[1865,2007] and ($B^\pm , B^*$)[5279,5325] (experimental values 
of the meson masses within the square brackets in MeV). The ground state
masses of the pseudoscalar  mesons comes from the solution of 
Eq.(\ref{trenv23}) with  
$A(k)$ given by Eq.(\ref{phsp}). The pion mass is fixed to its experimental 
value and the mass $m_2$ of one of the constituents quarks are varied with 
$m_1=$ 406 MeV and $\alpha$ constant. We choosed the values for 
$\alpha=$ 0.18, 0.4 and 0.5. 

The vector meson mass is associated to
the sum $m_1+m_2$. In that sense, the vector meson mass does not have 
the contribution of the strong attractive hyperfine interaction, 
and the Coulomb effective
attraction produces a  binding energy too small compared to the mass, 
which we have desconsidered here.  
In figure 3, the results 
of the difference of the squared masses $m^2_v-m^2_{ps}$ 
of the ground state of the vector mesons and pseudo-scalar mesons
as a function of the mass of the ground state of the pseudo-scalar
mesons $m_{ps}$ are shown. The experimental results are reasonably
described with $\alpha=0.4$. The value of $\alpha$ used to describe
the data depends on $m_1$. In our previous model calculation,
the value of 386 MeV was used and $\alpha$ was found to be 0.5 \cite{tp}. 
 The linear raising behaviour of
the difference $m^2_v-m^2_{ps}$ with $m_{ps}$ observed
in figure 3, is due to the  saturation of the binding energy when
$m_2\rightarrow \infty$. It is not clear if a confining potential, not
present in the model, can change such a trend.

\begin{figure}[bhpt] 
\begin{center}\scalebox{0.4}{\includegraphics{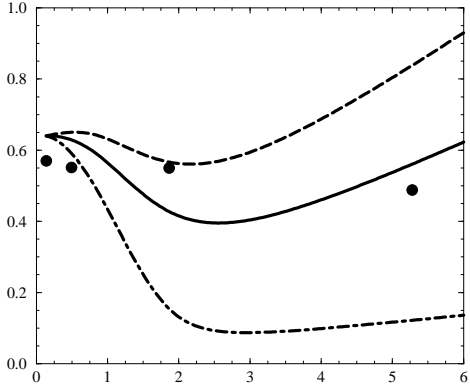}}
\end{center}
\vspace{-2 cm}
\caption{\label{fig3}
The difference $m^2_v-m^2_{ps}$(GeV$^2$) is plotted  
as a function of $m_{ps}$(GeV). Results from
Eq.(\protect{\ref{trenv23}}), with $\alpha=0.18$ (dot-dashed curve),
$\alpha=0.4$ (solid curve) and $\alpha=0.5$ (dashed curve). 
Experimental values  (solid circles).
}\end{figure}

\section{Conclusion}

The ``subtraction method'' \cite{t1} was applied to renormalize the 
$\uparrow\downarrow$-model\cite{pauli4} 
which contains an effective Coulomb interaction and
a hyperfine zero-ranged singular. We  have compared with
a different renormalization scheme that make use of regularization 
and subsequent renormalization\cite{pauli4}.  
The two drastically different 
schemes, both conceptually and numerically,  agree. 
Here we provide one more simple example, that the physics of the 
renormalized theory does not recognize the 
intermediate steps one performs to  mathematically 
define the initial undefined theory. 

{\bf Acknowledgments:} TF thanks to H. Leutwyler for
 suggesting the plot of figure 3.
TF also thanks to H.C. Pauli for the warm hospitality at the 
Max-Planck Institute in Heidelberg, where this work has 
been written, and to CNPq and FAPESP for financial support.

\appendix

\section{Equations in Momentum Space}
The effective model of \cite{pauli4} corresponds to use
the non-relativistic phase-space $A(k)=1$ in Eq.(\ref{mass2}) and a 
smeared delta-interaction of a Yukawa form:
\begin{eqnarray}
m^2_\pi\varphi (\vec k)=
\left[ 4 m^2+4k^2\right]\varphi(\vec k)-
\frac{4}{3\pi^2}\alpha
\int \frac{d\vec p}{m}
\nonumber \\
\times \left(\frac{2m^2}{(\vec k -\vec{p})^2}+
\frac{\eta^2}{\eta^2+(\vec k -\vec{p})^2}\right)
\varphi (\vec p) .
\label{mass1c}
\end{eqnarray}

In the renormalized model for the Coulomb plus Dirac-delta
interaction the bound state masses of 
the pion ground and excited states in s-wave,
are found numerically from the zeroes of Eq.(\ref{trenv23}): 
\begin{eqnarray}
\int^\infty_0 dp \frac{4\pi p^2}{A(p)} \left[\frac1{{m^*_\pi}^2-M_0^2(p)}
-\frac1{m_\pi^2-M_0^2(p)}
\right]
\nonumber \\
 +8\pi^2\int^\infty_0 dq \frac{q^2}{\sqrt{A(q)}} 
\int^\infty_0 dp \frac{p^2}{\sqrt{A(p)}}\;\;\;\;\;\;\;\;\;\;
\nonumber \\
 \times\left[\frac{t^V(p,q;{m^*_\pi}^2)}
{({m^*_\pi}^2 - M_0^2(p))({m^*_\pi}^2-M_0^2(p))} \right. \;\;\;\;\;\;\;\;\;
\nonumber \\ 
- \left. \frac{t^V(p,q;m^2_\pi)}{(m^2_\pi - M_0^2(p))
(m^2_\pi - M_0^2(q))}
\right]
=0,\;\;\;\;\; 
\label{p23r}
\end{eqnarray} 
with $m^*_\pi$ the excited s-wave state mass. The free mass of the
two quark system is $M_0(k)=\sqrt{k^2+m_1^2}+\sqrt{k^2+m_2^2}$.
The s-wave projected T-matrix of the Coulomb potential
 in Eq.(\ref{p23r}) is 
\begin{eqnarray}
t^V(p,q;M^2)=\int^1_{-1}dcos(\theta)<\vec p|T^V(M^2)|\vec q> ; \!\!
\label{t23sw}
\end{eqnarray} 
which is the solution of
\begin{eqnarray}
t^V(p,q;M^2)=\frac{4m}{3\pi^2}\frac{\alpha}{pq}
\frac{\ln\frac{(p-q)^2}{(p+q)^2}}{\sqrt{A(p)A(q)}}\;\;\;\;\;\;\;\;
\nonumber \\
+\frac{8m}{3\pi}\alpha\int^\infty_0dp'\frac{p'\ln\frac{(p-p')^2}{(p+p')^2}}
{p\sqrt{A(p)A(p')}}
\frac{t^V(p',q;M^2)}{M^2-M_0^2(p)}  , 
\label{t23r}
\end{eqnarray} 
the momentum space representation of s-wave projection of the
scattering equation for the T-matrix $T^V(M^2)$.


\begin{thebibliography}{99}

\bibitem{t1} T. Frederico, A. Delfino, and  L. Tomio,
Phys. Lett. {\bf B481} (2000) 143; T. Frederico, A.Delfino, L.Tomio, 
and V.S. Tim\'oteo,
"Fixed Point Hamiltonians in Quantum Mechanics", 
hep-ph/0101065.


\bibitem{pauli4} H.C. Pauli,
in: 
New Directions
in Quantum Chromodynamics, C.R. Ji and D.P. Min, Eds., AIP (1999) 80-139;
Nucl. Phys. {\bf B} (Proc. Supp.) {\bf 90} (2000) 154; ibid., 259.
 
\bibitem{glazek}S. D. Glasek and K. G. Wilson, \\
Phys. Rev. {\bf D57} (1998) 3558.

\bibitem{tp} T. Frederico and H.C. Pauli,\\ Phys. Rev. {\bf D64} (2001) 054007

\bibitem{frewer} M. Frewer, T.Frederico and H.C. Pauli, \\ these Proceedings.




\end{thebibliography}
\end{document}